\documentclass[twocolumn,graphicx,amsmath]{revtex4}
\usepackage{graphicx,amsmath,amsfonts,amssymb}
\newcommand{\BTO}{BaTiO$_3$}
\newcommand{\PTO}{PbTiO$_3$}
\newcommand{\STO}{SrTiO$_3$}

\draft 

\begin{document}


\title{High field properties of typical perovskite ferroelectrics by first-principles modeling} 



\author{Daniel I. Bilc}
\email[]{Daniel.Bilc@itim-cj.ro}
\author{Liviu Zarbo} 
\author{Sorina Garabagiu}
\affiliation{Mol $\&$ Biomol Phys Dept, Natl Inst Res $\&$ Dev Isotop $\&$ Mol Technol, RO-400293 Cluj-Napoca, Romania}
\author{Eric Bousquet}
\affiliation{Physique Th\'eorique des Mat\'eriaux, Universit\'e de Li\`ege (B5), B-4000 Li\`ege, Belgium}
\author{Liliana Mitoseriu}
\email[]{lmtsr@uaic.ro}
\affiliation{Department of Physics, Alexandru I. Cuza University, RO-700506 Iasi, Romania}


\date{\today}

\begin{abstract}
Using first-principles calculations, we estimated the impact of large applied electric $E$ fields on the structural, dielectric, and ferroelectric properties of typical ferroelectrics.  
At large fields, the structural parameters change significantly, decreasing the strain between the different structural phases. This effect favours a polarization rotation model for ferroelectric switching in which the electronic polarization rotates between the directions of tetragonal, rhombohedral and orthorhombic phases. We estimate coercive fields $E_c$$\sim$31 MV/m and $\sim$52 MV/m at zero temperature for bulk ferroelectric monodomains of \BTO\ and \PTO, respectively. The dielectric permittivity and tunability of \BTO\ are the least affected at large fields, making this material attractive for applications in electronics and energy storage.
\end{abstract}

\pacs{71.15.-m, 71.15.Mb, 77.80.Fm, 77.84.-s, 85.50.-n, 88.85.jp}

\maketitle 


Understanding the behaviour of advanced oxide materials under applied finite electric field $E$ is very important for high performance technological applications. Many of such tunable oxide materials are composites with various nano/microstructural characteristics designed between typical ferroelectric (FE), like BaTiO$_3$, PbTiO$_3$ and \STO, and linear dielectric materials.  In such compounds,  the electric field $E$ can vary strongly at local level, and affect the local structural, dielectric and FE properties. Such properties as electrostriction, tunability, dielectric loss, polarization switching are of great interest and they enter into the coefficient of performance of many devices designed for advanced technological applications. In terms of wireless communication technologies, the tunability is one of the essential characteristics which must be well characterized and optimized at high values in these composite ceramics.
The high interest  in  such oxide materials is also motivated by the more recent finding of coexistence of ferroelectricity with the metallic character of surface states in topological FE oxides, which opens new perspectives for electronic, spintronic and quantum technology applications.\cite{Felser, Kee, Kim, Nagaosa}

In this respect, we studied the structural (unit cell volume and structural parameters), dielectric (dielectric permittivity $\epsilon$, and tunability), and FE (electronic polarization $P$ and polarization switching) properties of a few typical perovskite FE's under finite applied $E$ using first principles methods based on density functional theory (DFT), which goes beyond the more common approach based on empirical models such Landau theory~\cite{Landau, Ginzburg2, Devonshire1, Devonshire3} or Johnson model.\cite{Johnson}  Different first principles methods have been proposed to perform calculations under finite applied $E$~\cite{Fernandez, Umari, Cohen, Sai, Souza, Bellaiche, Dieguez} or electric displacement $D$~\cite{Hong, Hong2013} fields.
The practical calculations for periodic crystalline materials are more difficult to carry out than the zero field DFT calculations because the electric potential generated by $E$ field is linear, non-periodic and not compatible with the periodic boundary conditions. Moreover, the electric potential is unbounded giving rise to the Zener effect (interband tunneling). Thus, we have used the method developed by Vanderbilt {\it et. al.},\cite{Souza} which minimizes the electrical enthalpy functional with respect to Bloch type wave functions $\psi_{nk} (r) = e^{ikr} u_{nk} (r)$:
\begin{equation}
\label{Eq1}
\mathcal{F}[u_{nk};E]=U_{KS} - \Omega PE
\end{equation}
where $U_{KS}$ are the Kohn Sham energies  routinely determined in DFT, $- \Omega PE$ is the coupling energy between polarization $P$ and  $E$ field, $\Omega$ is the unit cell volume, and $u_{nk}$ are the Bloch wave functions. The two terms from electrical enthalpy are calculated considering the wave function $\psi_{nk}$ Bloch polarizable under the influence of external $E$ field. This method can be employed to perform atomic and structural relaxations in the presence of $E$ field for the materials of interest, and to estimate the field-dependent properties.

Our theoretical estimations of field-induced properties at zero absolute temperature were performed using the GGA approximation of Wu and Cohen, GGA-WC,\cite{GGA-WC} for the exchange and correlation energy functional as implemented in ABINIT code,\cite{ABINIT} and using optimized pseudopotentials generated with OPIUM.\cite{OPIUM}  The electronic states considered as valence states are: 4$s$, 4$p$, and 5$s$ for Sr, 5$s$, 5$p$, and 6$s$ for Ba, 6$s$, 6$p$, and 5$d$ for Pb,  3$s$, 3$p$, and 3$d$ for Ti, and 2$s$ and 2$p$ for O. An energy cutoff of 45 hartree was used for the plane-wave expansion of the wave functions and the Brillouin zone integrations were performed using 14$\times$14$\times$14 and 14$\times$12$\times$12 grids of $k$ points for five atom cubic/tetragonal/rhombohedral, and ten atom orthorhombic structures, respectively. The self-consistent-field calculations were considered to be converged when the total energy changes were smaller than 10$^{-12}$ hartree. 
Atomic relaxations at fixed cubic lattice parameter for the cubic structures, and full relaxations (atomic and cell geometry) for the tetragonal, rhombohedral, and orthorhombic structures were performed until the forces were smaller than 5$\times$10$^{-7}$ hartree/bohr.

GGA-WC describes very well the structural and FE properties of typical ferroelectrics and does not have the supertetragonality problem (overestimation of $c/a$ ratio, atomic displacements and $P$) which usual GGA's have for the tetragonal structures.\cite{Bilc2008} The values of quantities describing the spontaneous structural and FE properties (for  $E=0$) for the tetragonal structures of BaTiO$_3$ and PbTiO$_3$ are included in Table~\ref{Table1}. The overestimation of spontaneous tetragonality $c/a$ ratio and polarization $P_s$ for BaTiO$_3$ are due to the accuracy of GGA-WC pseudopotentials used in these calculations. Note that our previous estimations of the structural and FE properties for BaTiO$_3$ and PbTiO$_3$ using the same GGA-WC functional and CRYSTAL code are in a better agreement with the experimental values.\cite{Bilc2008}
\begin{table}[t]
\caption{\label{Table1}Optimized $a$ lattice constant, tetragonality $c/a$, atomic displacements $dz$ in fractions of c lattice constant, and macroscopic spontaneous polarization $P_s$ for tetragonal structures of \BTO\ and \PTO\ within GGA-WC functional. The experimental values are also given.  }
 \begin{tabular*}{0.48\textwidth}%
    {@{\extracolsep{\fill}}ccccc}
\hline\hline
                & \multicolumn{2}{c}{BaTiO$_3$} &  \multicolumn{2}{c}{PbTiO$_3$} \\
                & GGA-WC & Exp. &  GGA-WC & Exp. \\ 
\hline
$a$(\AA) & 3.973 & 3.986$^a$  & 3.89  & 3.88$^b$ \\
$c/a    $ & 1.023 & 1.010$^a$ & 1.093  & 1.071$^b$ \\
$dz_{\rm Ti}$ & 0.015 & 0.015$^a$  & 0.038 & 0.040$^c$ \\
$dz_{\rm O_{\parallel}}$ & -0.03 & -0.023$^a$  & 0.119 & 0.112$^c$ \\
$dz_{\rm O_{\perp}}$ & -0.019 & -0.014$^a$ & 0.124 & 0.112$^c$ \\
$P$(C/m$^2$) & 0.34 & 0.27$^d$ &  0.97  &0.5-1.0$^e$ \\ 
\hline\hline
\multicolumn{5}{l}{ $^a$Room temperature data from Ref.~\cite{Shirane1957}. } \\
\multicolumn{5}{l}{ $^b$Extrapolated to 0 K data from Ref.~\cite{Mabud1979}.  }\\
\multicolumn{5}{l}{ $^c$Room temperature data from Ref.~\cite{Shirane1956}.  }\\
\multicolumn{5}{l}{ $^d$Ref.~\cite{Wieder1955}.  } \\ 
\multicolumn{5}{l}{ $^e$Ref.~\cite{Lines1977}.  } \\
\end{tabular*}
\end{table}    

\begin{figure}[t]
\centering\includegraphics[scale=0.25]{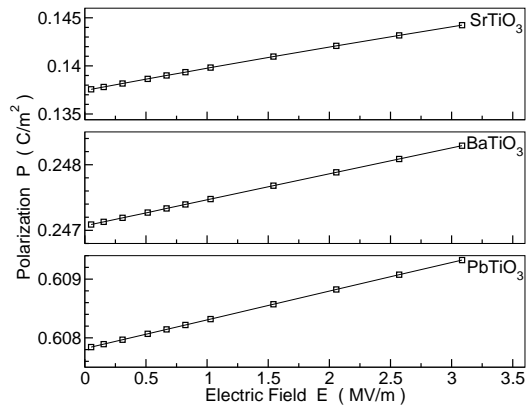}\\[-10pt]
\caption{\label{PscFig} (Color online) Electronic polarization $P$ dependence on electric $E$ field for the bulk cubic perovskite structure of \STO, \BTO\ and \PTO.}
\end{figure}

The next step was to estimate the field-dependence of polarization $P(E)$ and of static dielectric permittivity $\epsilon_s = \epsilon(E)$ for the cubic structures of \STO, \BTO\ and \PTO. The cubic lattice parameter was kept constant, but the electronic and atomic positions have been relaxed in the presence of $E$ field, which was oriented in the crystallographic direction $z$. We followed this procedure to estimate $P(E)$ and $\epsilon(E)$, which correspond to the region of transition from the cubic to tetragonal structure in the phase vs temperature diagram, where the values of $\epsilon(E)$ are maximized at the transition zone boundary. The polarization $P(E)$ dependence on field is shown in Figure~\ref{PscFig}. 
In order to estimate $\epsilon(E\rightarrow0)$ for small fields, we considered the dependence of $P(E)$ on the dielectric susceptibilities including up to the cubic terms, $P(E) = \chi^{(1)}E + \chi^{(2)}E^{2} + \chi^{(3)}E^{3}$, and  $\epsilon(E) = 1 + \chi^{(1)}(E)/\epsilon_0$, where $\chi^{(1)}(E)$ is the linear term of susceptibility and $\epsilon_0$ is the vacuum permittivity.  By fitting $P(E)$ values for small $E$ fields, we estimate $\epsilon_s(E\rightarrow0)$ values of: $\sim$270, $\sim$46 and $\sim$56 for \STO, \BTO\ and \PTO, respectively. The $\epsilon_s$ value of $\sim$270 for  \STO\ is smaller than the value of 391 estimated within LDA from Ref.~[\onlinecite{Antons}]. These theoretical values of $\epsilon_s$ for \STO\  compare well with the room temperature experimental value of $\sim$300, but they are largely underestimating the zero temperature experimental value of $\sim$20000.\cite{Sakudo} At low temperature the antiferrodistortive (AFD) oxygen octahedral rotations (a$^0$a$^0$c$^-$ in Glazer notations~\cite{Glazer}) are present in \STO, which strongly influences the polar ferroelectric mode and $\epsilon_s$. The compressive inplane strain to \STO, increases these AFD rotations and $\epsilon_s$ which is comparable to the zero temperature experimental value for strain corresponding to transition zone boundary between paraelectric and FE tetragonal structures.\cite{Antons} The estimation of $\chi^{(1)}(E)$ at a given $E$ field can be made by the method of finite differences $\chi^{(1)}(E) = \Delta P(E)/\Delta E$. Using this method, we obtained the following $\epsilon_{s}(E)$ values of $\sim$235, $\sim$46, and $\sim$56 at 3 MV/m for \STO, \BTO\ and \PTO, respectively. These results show that \STO\ has a large relative tunability $n_{r}=[\epsilon_s(E\rightarrow0) - \epsilon_s(E)]/\epsilon_s(E\rightarrow0)$ of $\sim$ 13$\%$ at 3 MV/m,  $n_{r}$ value that increases strongly at transition zone boundary,\cite{Antons} whereas the cubic phases of \BTO\ and \PTO\  do not show any tunability.

\begin{figure}[h]
\centering\includegraphics[scale=0.3]{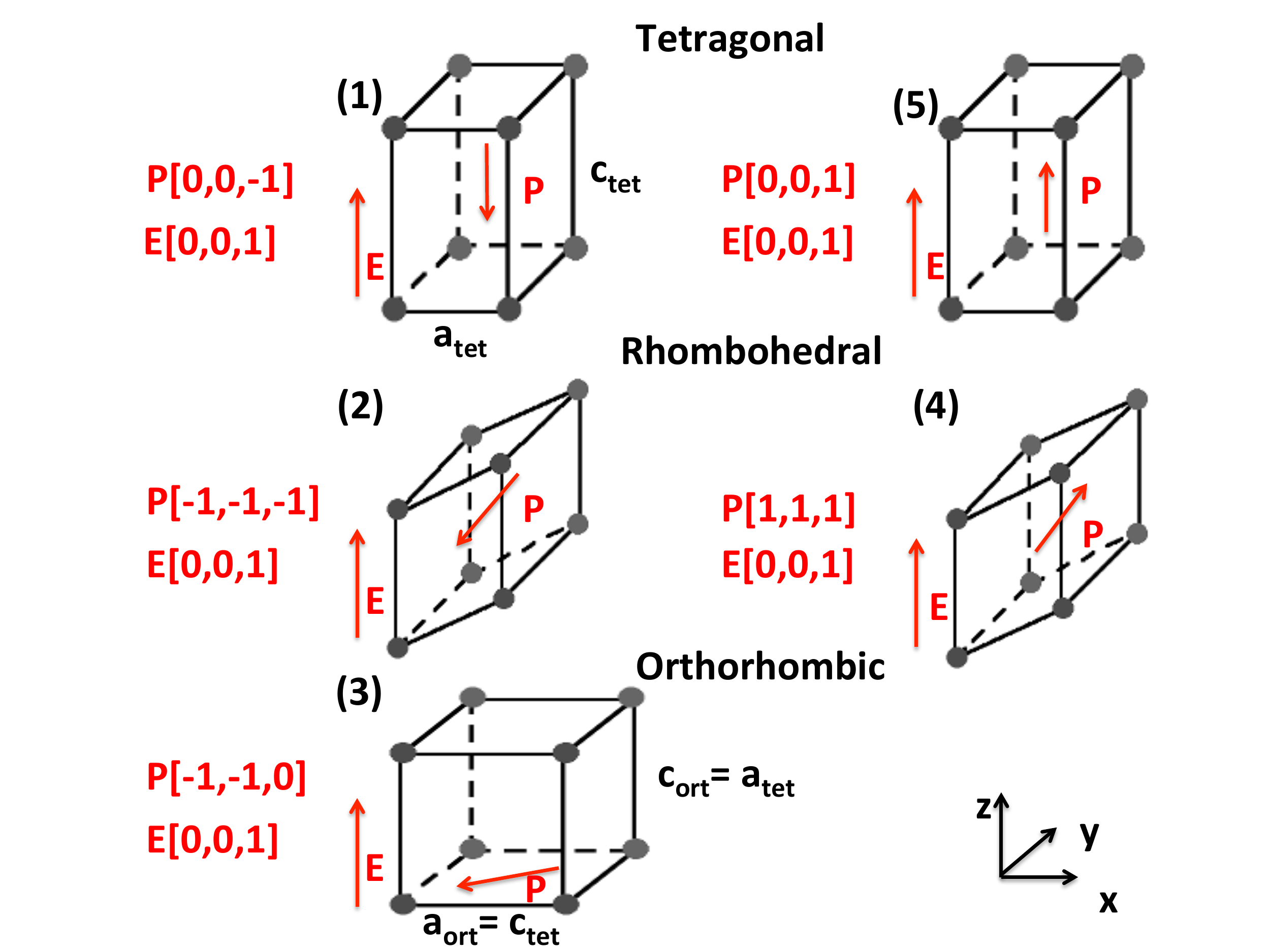}\\[-10pt]
\caption{\label{SwitchModel} (Color online) The switching model through rotation of polarization $P$ in external $E$ field. The switching is achieved by $P$ rotation between the different phases: tetragonal (1) and (5), rhombohedral (2) and (4), and orthorhombic (3). $E$ field is oriented in [0,0,1] direction ($z$) and $P$ along the directions given by structural phases.}
\end{figure}

For modeling the FE switching properties, we considered a model in which polarization rotates between the different polarization directions of the structural phases: tetragonal, orthorhombic,  and  rhombohedral, as the strength of applied field increases in a given direction (Figure~\ref{SwitchModel}). This approach was not often used  in literature because the Ising model is believed to be favoured by the strain conditions   to which the switching units ({\it histerons}) are subjected. Ising model involves the crossing of switching units through the high-symmetry structures (paraelectric structures) and the cancellation of polarization, $P = 0$. Previous theoretical studies for \BTO\  showed that field-induced polarization rotation can occur,\cite{Garcia} and is responsible for the ultrahigh electromechanical response.\cite{Cohen} High piezoelectric effects and field-induced transition in \BTO\ single-crystals were explained by a coherent polarization rotation within a Landau-based approach combined with crystalline anisotropy and field effects.\cite{Zang} Close to the structural FE-FE transitions of \BTO, the polarization rotation instability was considered as responsible for the divergence of shear elastic compliance.\cite{Nanni} 

\begin{figure}[t]
\centering\includegraphics[scale=0.25]{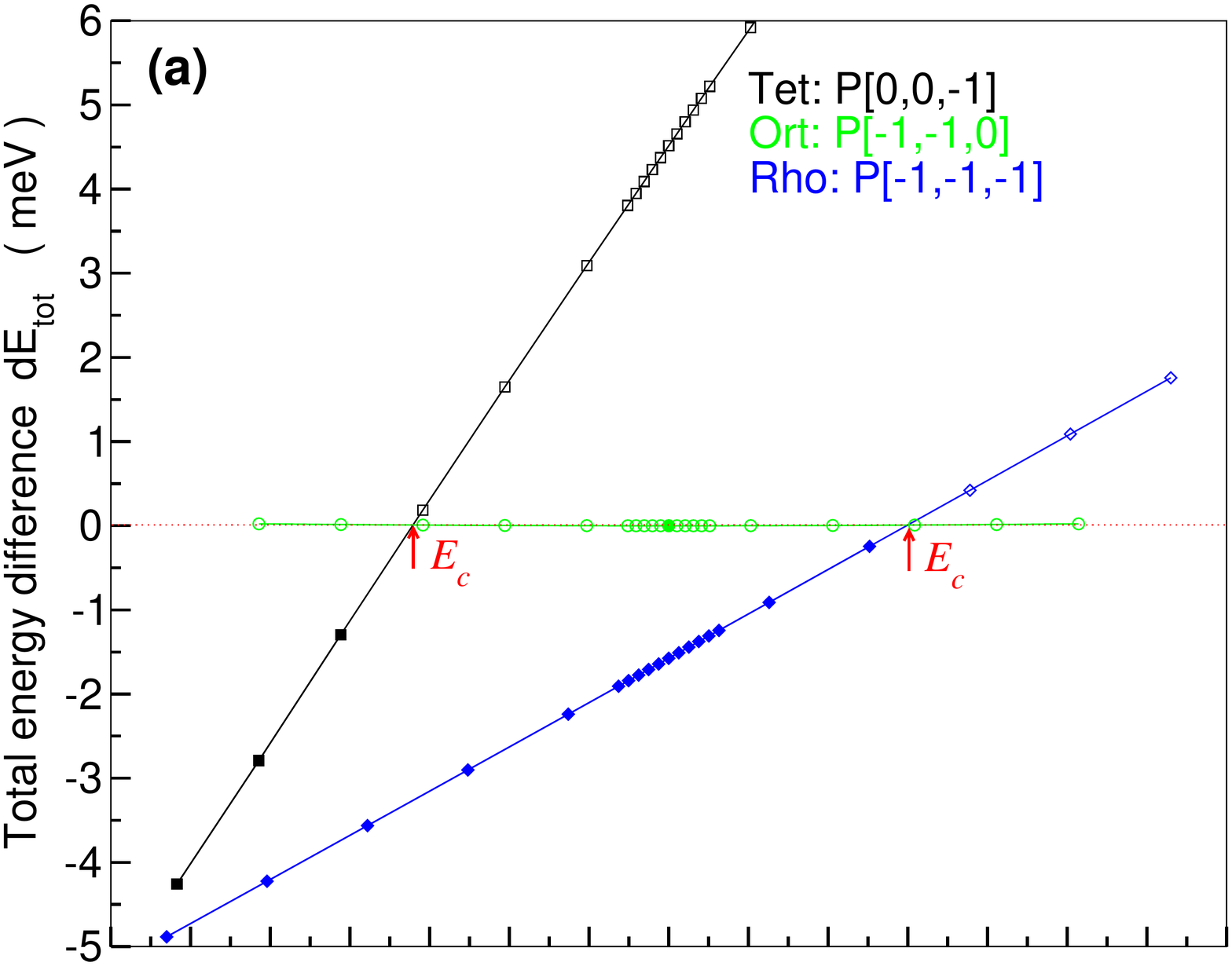}\\
\centering\includegraphics[scale=0.25]{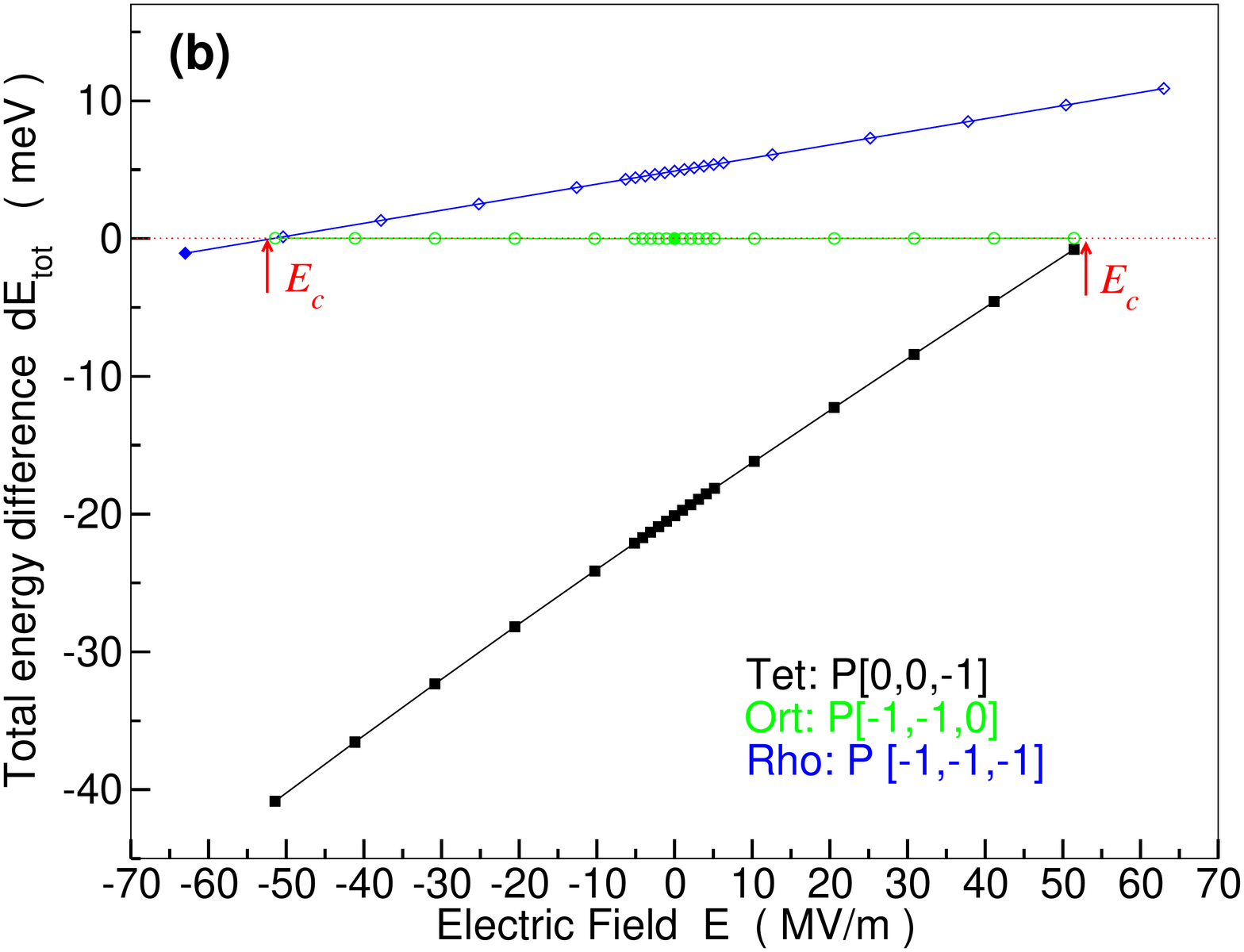}\\[-10pt]
\caption{\label{EnFig} (Color online) Electric field dependence of the total energy difference $d$E$_{tot}$ relative to orthorhombic structure for the different structural phases of a bulk monodomain:  (a) \BTO\  and (b) \PTO. Negative $d$E$_{tot} <$ 0 values favourable in terms of structural stability are represented by filled symbols. Coercive $E_c$ fields, where all structural phases have comparable total energies are marked by arrows. The positive $E$ fields are oriented along [0,0,1] direction.}
\end{figure}

We obtain the coercive fields $E_c$ from the dependence on $E$ of total energy corresponding to electric enthalpy given by Eq.~\ref{Eq1} for the different structural phases (see Figure~\ref{SwitchModel}), which were fully relaxed (atomic and stress relaxations) in the presence of field. In Figure~\ref{EnFig}(a),  we show this dependence for the total energy difference $d$E$_{tot}$ of a \BTO\  monodomain relative to the orthorhombic structure for which $E \perp P$ (Figure~\ref{SwitchModel}(3)). The orthorhombic structure is chosen as a reference for total energy since it is the least affected by $E$ field as $\Omega PE$ term is negligible. Structures with $d$E$_{tot}$ $<$ 0, are favourable in terms of structural stability (filled symbols in Figure~\ref{EnFig}(a)). 
The coercive fields $E_c$ were estimated from the field values at which the three structural phases (tetragonal, rhombohedral and orthorhombic) are comparable in energy ($d$E$_{tot}$ $\sim$ 0 meV) and $P$ can easily rotate between the directions corresponding to the three \BTO\ symmetries. These results suggest coercive fields $E_c \sim$ 31 MV/m for a \BTO\  monodomain. This $E_c$ value agrees very well with the field value of 30 MV/m at which the large piezoelectric response was estimated within LDA for \BTO~\cite{Cohen, Stolbov} and the values of $\sim$20 MV/m~\cite{Waghmare}  and $\sim$150 MV/m~\cite{Stolbov} computed by molecular dynamics simulations at 100 K.

\begin{figure}[t]
\centering\includegraphics[scale=0.25]{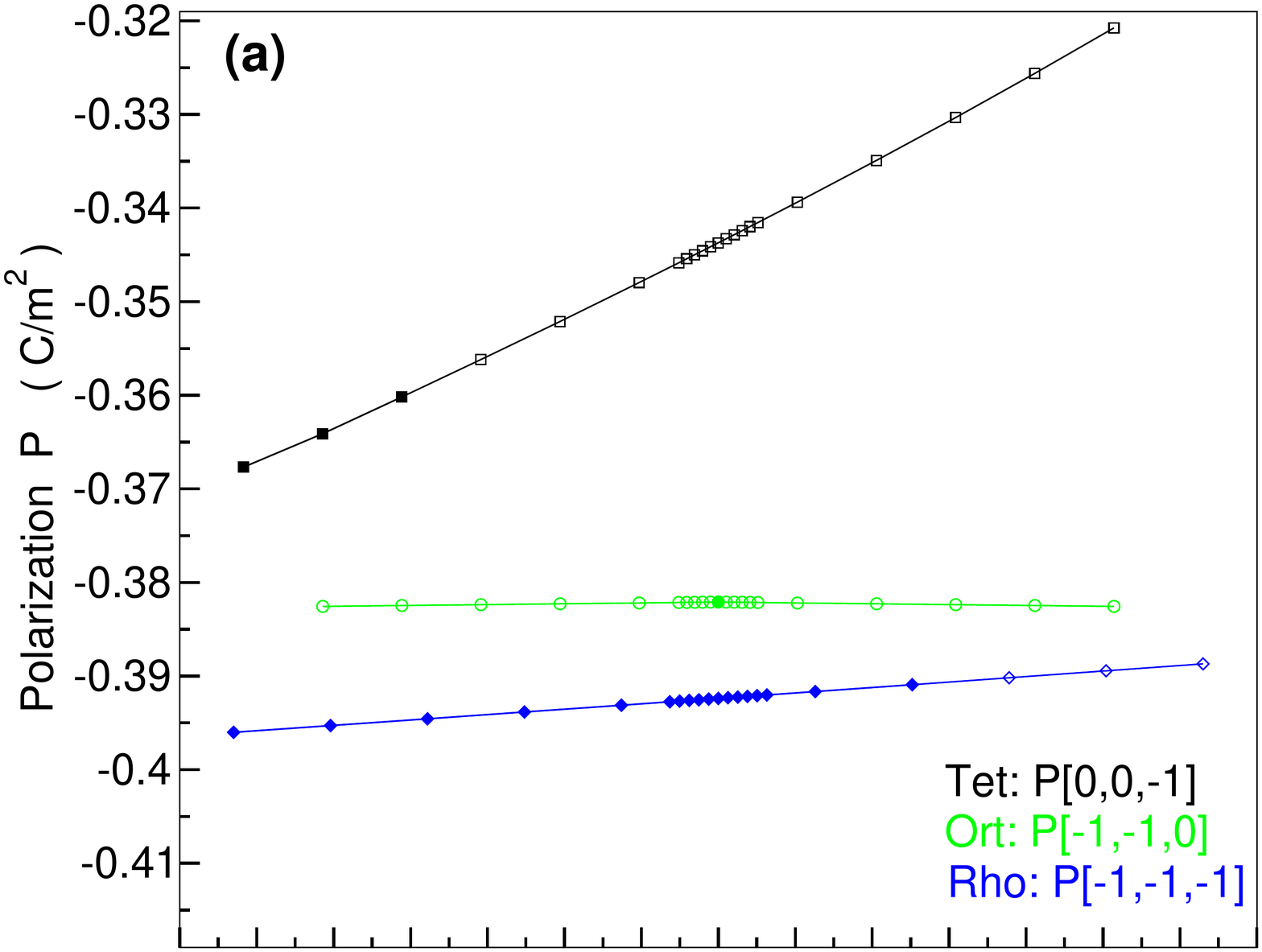}\\
\centering\includegraphics[scale=0.25]{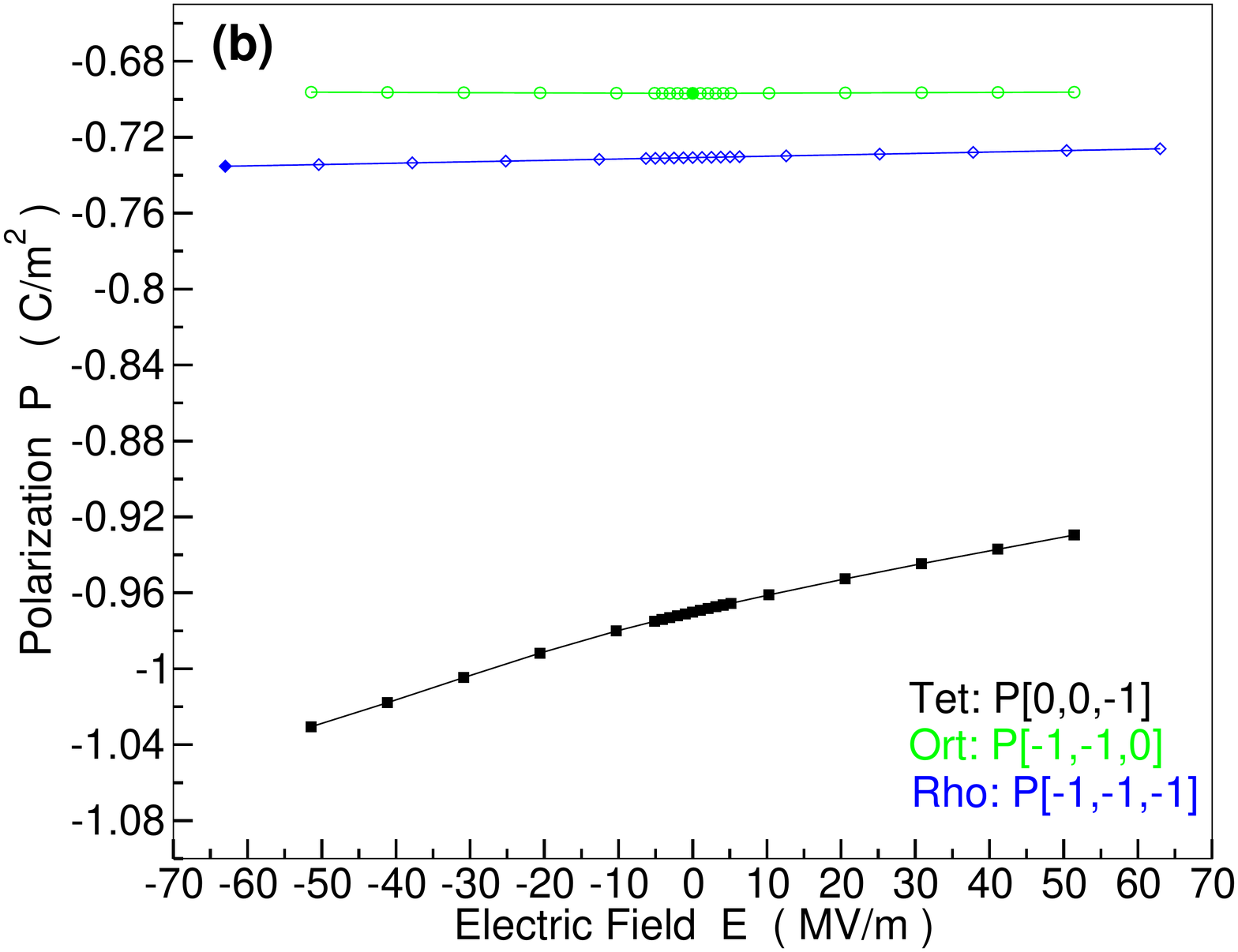}\\[-10pt]
\caption{\label{PFig} (Color online) Electronic polarization $P$ dependence on electric $E$ field for the different structural phases of a bulk monodomain:  (a) \BTO\  and (b) \PTO. $P$ values corresponding to $d$E$_{tot} <$ 0 are represented by filled symbols. The positive $E$ fields are oriented along [0,0,1] direction.}
\end{figure}

For the switching mechanism of a \PTO\  monodomain, which has the polar tetragonal ground structure, we considered also the other polar orthorhombic and rhombohedral structures as "metastable", similar to those of \BTO.  For $E$ values $\sim$52 MV/m, which are higher compared to those of \BTO, the total energy of the three phases are comparable ($d$E$_{tot}$ = 0, Figure~\ref{EnFig}(b)). The higher coercive field $E_c \sim$ 52 MV/m of  \PTO\  is due to larger $d$E$_{tot}$ values in the absence of field by comparison with the case of \BTO\  ($|d$E$_{tot}|$ $\sim$ 20 meV for \PTO\  and $|d$E$_{tot}|$ $\sim$ 5 meV for \BTO\  at $E$ = 0). This $E_c$ value correspond to a linear piezoelectric response of \PTO\ \cite{Roy}, being smaller than the value of $\sim$270 MV/m estimated by molecular dynamics simulations at 100 K.\cite{Cohen2011}
We would also like to emphasize that \PTO\  has double well energy, $d$E$_{dw}$ (total energy difference between polar tetragonal and nonpolar cubic structures) and a much higher polarization than those of  \BTO\  ($d$E$_{dw}$ $\sim$ 100 meV, $P$ $\sim$ 0.98 C/m$^2$ for  \PTO, and  $d$E$_{dw}$ $\sim$ 14 meV and $P$ $\sim$ 0.26 C/m$^2$ for \BTO).\cite{Bilc2008}  These results show that the required energy $d$E$_{tot}$ to be supplied externally by the electric field $E$ to a monodomain in the polarization rotation mechanism of switching are much lower compared with the necessary energy $d$E$_{dw}$ for the Ising model of polarization switching. The results suggest the polarization rotation as the dominant switching mechanism at high $E$ fields compared to the Ising model of switching. Our results are supported by the first principles DFT calculations performed for \PTO\ under finite $D$ field within LDA, which find the polarization rotation through tetragonal, rhombohedral and orthorhombic phases  to be more favourable than the Ising model of polarization switching that requires $E_c$ values of  $\sim$250 MV/m.\cite{Hong} 

\begin{figure}[t]
\centering\includegraphics[scale=0.19]{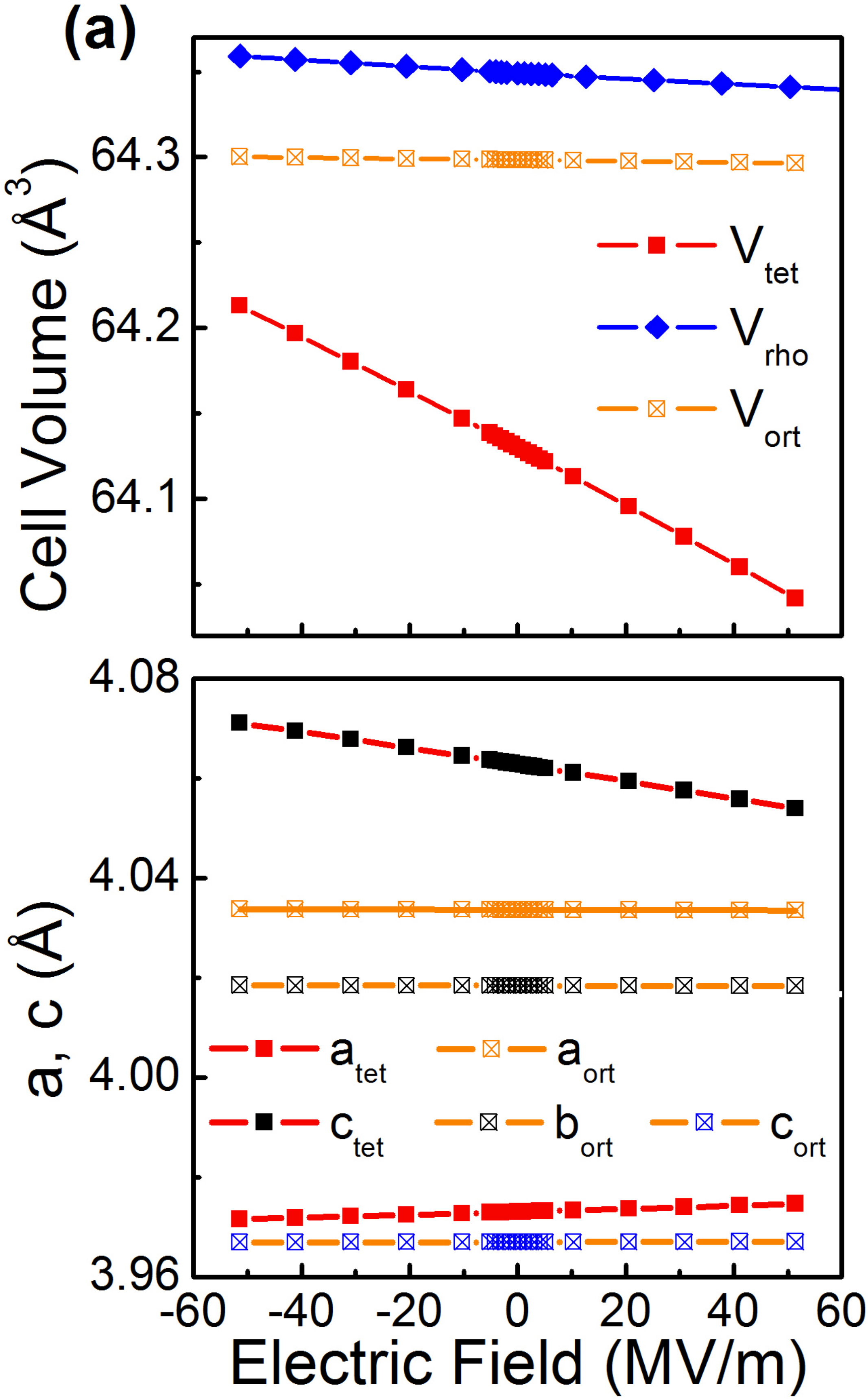}%
\centering\includegraphics[scale=0.19]{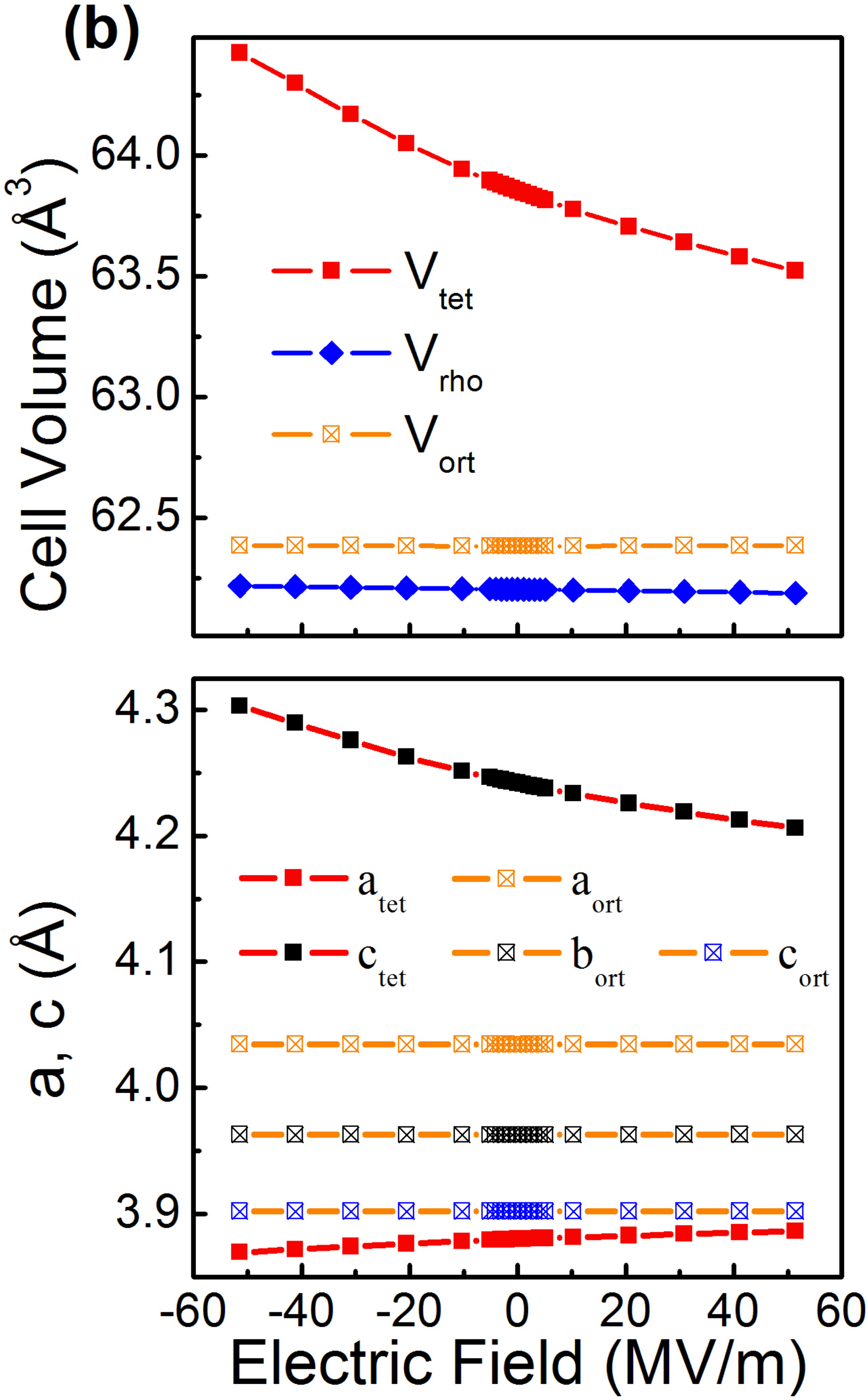}\\[-10pt]
\caption{\label{StFig} (Color online) Electric field dependence of the unit cell volume and structural parameters for the different phases of bulk:  (a) \BTO\  and (b) \PTO. }
\end{figure}

The structural properties are significantly affected by the applied field (Figure~\ref{StFig}). The unfavourable energetic case of the tetragonal structure in which $E>$ 0 and $E$ is antiparallel to $P$ (Figure~\ref{SwitchModel}(1)) leads to a decrease in the unit cell volume and $c_{tet}$ parameter such that the strain relative to $a_{ort}$ parameter of orthorhombic structure is reduced supporting  $P$ rotation through which $c_{tet}$ becomes $a_{ort}$ and $a_{tet}$ becomes $c_{ort}$. Similarly, the favourable energetic case in which $E <$ 0 and $E$ is parallel with $P$ (Figure~\ref{SwitchModel}(5)) enlarges the unit cell volume and $c_{tet}$ parameter, enlargements which are limited by the strain relative to $a_{ort}$ parameter.

\begin{figure}[t]
\centering\includegraphics[scale=0.25]{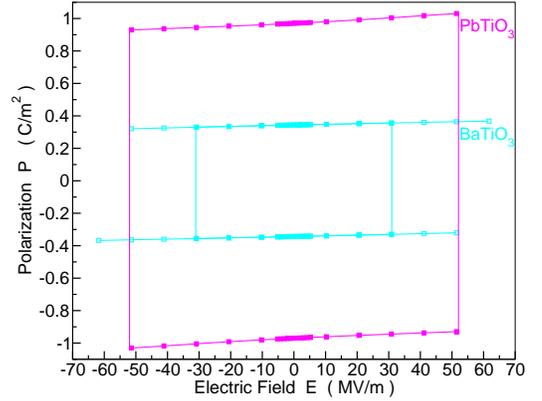}\\[-10pt]
\caption{\label{HisteronFig} (Color online) Histeron for the tetragonal phase of \BTO\  and \PTO\  bulk monodomains. Coercive $E_c$ fields are indicated by vertical lines. Polarization values corresponding to $d$E$_{tot} <$ 0 are represented by filled symbols.}
\end{figure}

At high $E$ fields ($E \sim E_c$), the polarization is also significantly affected. In the tetragonal structures, $P$ differs by $\sim 3\%$ for \BTO\  at $E$ = 31 MV/m and respectively by $\sim 6\%$ for \PTO\  at $E$ =  52 MV/m (Figure~\ref{PFig}(a) and (b)). Considering the field-induced polarization   variations from Figure~\ref{PFig} and the estimated coercivity $E_c$ values, we show in Figure~\ref{HisteronFig} the histeron model for tetragonal phase of \BTO\  and \PTO\ monodomains,  which may be further used in multiscale modeling methods required to describe the properties of complex ferroelectric materials. Fitting $P(E)$ values from Figure~\ref{PFig},  we estimate $\epsilon_s(E\rightarrow0)$ $\sim$48 and $\sim$110,  for the tetragonal phases of \BTO\  and \PTO, respectively. The finite difference method gives the following  $\epsilon_s(E)$ values of $\sim$49 and $\sim$104 at $E$ = 3 MV/m for the tetragonal FE phases of \BTO\  and \PTO, respectively.  These results show an enhanced dielectric tunability for the tetragonal FE phase of  \BTO\  and \PTO, comparing with that of the cubic paraelectric phase. 

In  this  work,  we  show  the role of large external fields on the  structural,  dielectric  and  FE  properties  of  a few typical ferroelectric perovskites.  Whenever present in various types of material combinations (solid solutions, composites, multilayers), the properties of ferroelectrics are strongly affected by electric fields and this may significantly affect the device performance.
We find \BTO\ to have comparable static dielectric permittivities in the cubic and tetragonal bulk phases, which combined with the high dielectric tunability of \STO, make (BaSr)TiO$_3$ solid solutions among the most interesting materials for tunability applications, since they exhibit slight variation of dielectric permittivity in a large range of temperatures. \BTO\ shows the smallest dielectric tunability, making it a good candidate to be used in composites for energy storage applications, for which large dielectric permittivity, small tunability and losses are require.  The FE monodomain states of \BTO\ and \PTO\ show high coercive $E_c$ fields of $\sim$30-50 MV/m.  These intrinsic $E_c$ fields fall between intrinsic values computed by other models or experiments considered to give intrinsic limits, since they were determined in ultrathin films such as: 5 MV/m in single crystal,\cite{Fridkins} 19.5 MV/m in Devonshire theory,\cite{Janovec} 70 MV/m in mean field theory,\cite{Fridkins} 25-160 MV/m in ultrathin polycrystalline films,\cite{Fridkins, Fridkins2, Jo} 196 MV/m in Ginzburg - Landau theory.\cite{Fridkins3}
The large electric fields favour a FE switching model in which electronic polarization rotates between the different polarization orientations of the structural phases. This model is supported by the continuous polarization switching mechanism (Bloch type switching), which was recently predicted to be favorable for bulk \BTO\  rhombohedral phase.\cite{Hlinka} We hope our findings will stimulate other more complex material studies under finite external $E$ field needed in many practical applications.


%
%

%

\begin{acknowledgments}

The authors acknowledge financial support from the Romanian National Authority for Scientific Research, CNCS-UEFISCDI, Project number PN-II-PT-PCCA-2013-4-1119. This work was supported by FRS-FNRS Belgium (E. B.) and E.B. acknowledges the Consortium des Equipements de Calcul Intensif (CECI), funded by the FRS-FNRS (Grant 2.5020.11).

\end{acknowledgments}


\begin{thebibliography}{99}

\bibitem{Felser} B. Yan, M. Jansen, and C. Felser, Nat. Phys. {\bf 9}, 709 (2013).
\bibitem{Kee} J. M. Carter, V. V. Shankar, M. A. Zeb, and H. Y. Kee, Phys. Rev. B {\bf 85}, 115105 (2012).
\bibitem{Kim} B. J. Yang and Y. B. Kim, Phys. Rev. B {\bf 82}, 085111 (2010).
\bibitem{Nagaosa} A. Shitade, H. Katsura, J. Kunes, X. L. Qi, S. C. Zhang, and N. Nagaosa, Phys. Rev. Lett. {\bf 102}, 256403 (2009).
\bibitem{Landau} L. D. Landau, I. M. Khalatnikov, Dokl. Acad. Nauk. SSSR {\bf 96}, 469 (1954).
\bibitem{Ginzburg2} V. L. Ginzburg, Sov. Phys. Solid State {\bf 2},1824 (1960).
\bibitem{Devonshire1} A. F. Devonshire, Phyl. Mag. {\bf 40}, 1040 (1949).
\bibitem{Devonshire3} A. F. Devonshire, Adv. Phys. {\bf 3}, 85 (1954).
\bibitem{Johnson} K. Johnson, J. Appl. Phys. {\bf 33}, 2826 (1962).
\bibitem{Fernandez} P. Fern\'andez, A. Dal Corso, A. Baldereschi, and F. Mauri, Phys. Rev. B {\bf 55}, R1909 (1997).
\bibitem{Umari} P. Umari and A. Pasquarello, Phys. Rev. Lett. {\bf 89}, 157602 (2000).
\bibitem{Cohen} H. Fu and R. E. Cohen, Nature {\bf 403}, 281 (2000).
\bibitem{Sai} N. Sai, K. M. Rabe, and D. Vanderbilt, Phys. Rev. B {\bf 66}, 104108 (2002).
\bibitem{Souza} I. Souza, J. \'I\~niguez, and D. Vanderbilt, Phys. Rev. Lett. {\bf 89}, 117602 (2002).
\bibitem{Bellaiche} H. Fu and L. Bellaiche, Phys. Rev. Lett. {\bf 91}, 057601 (2003).
\bibitem{Dieguez} O. Dieguez and D. Vanderbilt, Phys. Rev. Lett. {\bf 96}, 056401 (2006).
\bibitem{Hong} J. Hong and D. Vanderbilt,  Phys. Rev. B {\bf 84}, 115107 (2011).
\bibitem{Hong2013} J. Hong and D. Vanderbilt,  Phys. Rev. B {\bf 87}, 064104 (2013).
\bibitem{GGA-WC} Z. Wu and R. E. Cohen, Phys. Rev. {\bf B 73}, 235116 (2006).
\bibitem{ABINIT} X. Gonze, {\it et. al.}, Comput. Phys. Commun. {\bf 180}, 2582 (2009).
\bibitem{OPIUM} A. M. Rappe, K. M. Rabe, E. Kaxiras, and J. D. Joannopoulos, Phys. Rev. B {\bf 41}, 1227 (1990).
\bibitem{Bilc2008} D. I. Bilc, R. Orlando, R. Shaltaf, G. M. Rignanese, J. Iniguez, and Ph. Ghosez, Phys. Rev. B {\bf 77}, 165107 (2008).
\bibitem{Shirane1957} G. Shirane, H. Danner, and P. Pepinsky, Phys. Rev. {\bf 105}, 856 (1957).
\bibitem{Mabud1979}  S. A. Mabud and A. M. Glazer, J. Appl. Crystallogr. {\bf 12}, 49 (1979).
\bibitem{Shirane1956}  G. Shirane, P. Pepinsky, and B. C. Frazer,  Acta Cryst. {\bf 9}, 131 (1956).
\bibitem{Wieder1955}  H. H. Wieder, Phys. Rev. {\bf 99}, 1161 (1955).
\bibitem{Lines1977}  M. E. Lines and A. M. Glass { \it Principles and Applications of Ferroelectrics and Related Materials} ( Oxford University Press, Oxford, 1977), Chap.8.
\bibitem{Antons} A. Antons, J. B. Neaton, K. M. Rabe, and D. Vanderbilt,  Phys. Rev. B {\bf 71}, 024102 (2005).
\bibitem{Sakudo} H. Uwe and T. Sakudo, Phys. Rev. B {\bf 13}, 271 (1967).
\bibitem{Glazer} A. Glazer,  Acta Cryst. B.  {\bf 28}, 3384 (1972).
\bibitem{Garcia} A. Garcia, and D. Vanderbilt,  Appl. Phys. Lett. {\bf 72}, 2981 (1998).
\bibitem{Zang} H. Zang,  J. Appl. Phys. {\bf 113}, 184111 (2013).
\bibitem{Nanni} F. Cordero, H. T. Langhammer, T. M�ller, V. Buscaglia, and P. Nanni, Phys. Rev. B {\bf 93}, 064111 (2016).
\bibitem{Stolbov} S. Stolbov, H. Fu, R. E. Cohen, L. Bellaiche, and D. Vanderbilt. {\it  Fundamental Physics of Ferroelectrics 2000}, (R.E. Cohen, ed.) AIP Conference Proceedings 535, New York, pp. 151-158 (2000).
\bibitem{Waghmare} J. Paul, T. Nishimatsu, Y. Kawazoe, U.V. Waghmare, Phys. Rev. B {\bf 80}, 024107 (2009).
\bibitem{Roy} A. Roy, M. Stengel, and D. Vanderbilt, Phys. Rev. B  {\bf 81}, 014102 (2010).
\bibitem{Cohen2011} X. Zeng and R. E. Cohen, Appl. Phys. Lett.  {\bf 99}, 142902 (2011).
\bibitem{Fridkins} R. Gaynutdinov, M. Minnekaev, S. Mitko, A. Tolstikhina, A. Zenkevich, S. Ducharme, and V. Fridkin, Physica B, {\bf 424}, 8 (2013).
\bibitem{Janovec} V. Janovec, Czechosl. Journ. Phys. {\bf 8}, 3 (1958).
\bibitem{Fridkins2} R. Gaynutdinov, M. Minnekaev, S. Mitko, A. Tolstikhina, A. Zenkevich, S. Ducharme, and V. Fridkin, JETP Letters {\bf 98}, 339 (2013).
\bibitem{Jo} J. Y. Jo, Y. S. Kim, T. W. Noh, Jong-Gul Yoon, and T. K. Song, Appl. Phys. Lett. {\bf 89}, 232909 (2006).
\bibitem{Fridkins3}  V. M. Fridkin, and S. Ducharme, Phys. Solid State {\bf 43}, 1320 (2001).
\bibitem{Hlinka} M. Taherinejad, D. Vanderbilt, P. Marton, V. Stepkova, and J. Hlinka, Phys. Rev. B {\bf 86}, 155138 (2012).

\end{thebibliography}

\end{document}